\documentclass[11pt]{article}
\usepackage{times}
\usepackage{a4}
\usepackage{amsfonts}
\usepackage{amsthm}
\usepackage{amsmath}
\usepackage{eufrak}

\usepackage[all]{xy} 

\newbox\ncintdbox \newbox\ncinttbox 
\setbox0=\hbox{$-$}
\setbox2=\hbox{$\displaystyle\int$}
\setbox\ncintdbox=\hbox{\rlap{\hbox
    to \wd2{\hskip-.125em\box2\relax\hfil}}\box0\kern.1em}
\setbox0=\hbox{$\vcenter{\hrule width 4pt}$}
\setbox2=\hbox{$\textstyle\int$}
\setbox\ncinttbox=\hbox{\rlap{\hbox
    to \wd2{\hskip-.125em\box2\relax\hfil}}\box0\kern.1em}

\newcommand{\stroke}{\mathbin|}     

\def\proof{\noindent {\bf proof:\ }}
\def\endproof{\vrule height 0.5em depth 0.2em width 0.5em}
\newtheorem{lemma}{Lemma}[section]

\newtheorem{thm}[lemma]{Theorem}

\newtheorem{corollary}[lemma]{Corollary}
\theoremstyle{definition}

\newtheorem{defn}[lemma]{Definition}

\newtheorem{remark}[lemma]{remark}

\newcommand{\C}{\mathbb{C}}       
\newcommand{\R}{\mathbb{R}}       
\newcommand{\Z}{\mathbb{Z}}       
\newcommand{\0}{{\vphantom{\dagger}}} 

\newcommand{\bea}{\begin{eqnarray}}  
\newcommand{\eea}{\end{eqnarray}}  
\newcommand{\beas}{\begin{eqnarray*}} 
\newcommand{\eeas}{\end{eqnarray*}}  
\newcommand{\be}{\begin{equation}} 
\newcommand{\ee}{\end{equation}} 

\def\<#1,#2>{\langle#1\stroke#2\rangle} 

\def\Z{{\mathbb Z}}
\def\C{{\mathbb C}}
\def\R{{\mathbb R}}

\def\ei{\,\,\,{\rm l} \!\!\!\!\!\; 1\,}

\def\CA{{\cal A}} 
 
\def\CM{{\cal M}}

\def\CQ{{\cal Q}}
\def\CH{{\cal H}}

\def\oh{\frac{1}{2}}

\begin{document}
\begin{flushright} 
GA 2/2003
\end{flushright}  

\vspace*{3cm} 
\noindent

\begin{center} 
{\bf \Large Time evolutions in quantum mechanics  and (Lorentzian) geometry} 

\vspace{2cm} 

{\large  
Mario Paschke \footnote{E-mail: Mario.Paschke@math.slu.cz}\footnote{ \tt Alexander von Humboldt
research fellow} }

\vspace{1cm}

{\large \it Silesian University \\ Matematical Institute, \\ Opava, Czech
Republic }

\end{center} 

\vspace{3cm} 

\begin{abstract}
\noindent
As R.Feynman has shown to F. Dyson --  who published it then in 1990 under the name of "Feynman's
proof of Maxwell's equations"  -- the only interactions compatible
with the canonical uncertainty relation (for scalar particles
on flat $\R^3$) are the Lorentz covariant electromagnetic interactions.\\
We generalize Feynman's argument to arbitrary configuration spaces, thereby
clarifying its hidden assumptions as well as its geometrical significance. 
In addition, our result establishes a correspondence between globally hyperbolic spacetimes and
solutions to our (algebraically formulated) axioms for nonrelativistic quantum mechanics. 
\end{abstract}

\cleardoublepage
\section{Introduction}
It is a well known fact that the canonical uncertainty relation (cure), when written in the
form 
\begin{equation} 
m [x_k, \dot{x}_l ] = i\,\, \delta_{kl} \,\, ,  \label{heiss}
\end{equation}
i.e. using the velocities rather than the canonical momenta ,
provides severe restrictions on the admissible time evolutions of quantum mechanical
systems. 
If this time evolution respects Newton's second law $\ddot{x}_k =
F_k(\vec{x},\dot{\vec x}, t)$ , then it is always generated by a Hamiltonian $H$ which is
necessarily of the well-known form 
\[  H = \sum\limits_k \frac{1}{2m} \left(-i\frac{\partial}{\partial x_k}  - e
         A_k(\vec{x},t)\right)^2 +e\varphi(\vec{x},t)  \]
thus implying the existence of electric and magnetic fields which obey the homogeneous 
Maxwell equations. No other interactions (apart from the electromagnetic ones) are allowed.

\noindent
This result  has first
been published by F. Dyson \cite{Dyson} under the name of "Feynman's proof of Maxwell's equations". 
Dyson found it very remarkable, that
the {\em Lorentz-covariant} Maxwell equations can be derived starting from the {\em
nonrelativistic} uncertainty relation.

\noindent
In the following years, Feynman's result has then been generalized in various directions (see
\cite{Fein-revi} ), thereby also reproducing the interactions with the gravitational field \cite{Tani} and
(non-abelian) gauge fields \cite{Lee,Tani}. Other papers tried (without much success) to develop a
relativistic generalization of Feynman's argument.

\noindent
However, most of these authors have investigated classical systems, -- replacing the
commutator  by the symplectic form -- and in particular the question under which additional
conditions a dynamical system on a symplectic manifold can be cast into a Hamiltonian form.
Moreover they always used special coordinate systems, just as Feynman did. Thus, the generalization
to generic configuration spaces remained an open problem. Even more so, it has never been
investigated why  only {\em "geometric", Lorentz covariant} interactions, i.e.
gravitational and gauge interactions, are consistent with the uncertainty principle.

\noindent
We should point out that Feynman's proof (in quantum mechanics) relies on several hidden
assumptions. For instance, he uses that every operator which commutes with all the functions of
the $x_i$ must be a function of the $x_i$ itself, and, more importantly, the existence of a
faithful representation of the translation group and the algebra of functions itself. By the
Stone-von Neumann Theorem the Heisenberg-algebra ($\ref{heiss}$) possesses up to unitarity
equivalence only one representation with these properties. Thus, if
one thinks of the operators $(x_i(t), \dot{x}_i(t))$ as generating one such representation at each
time $t$, all these representations are unitarily equivalent. This establishes the
existence of unitary time evolution operators.
Using another well known theorem by Stone, one then immediately obtains the existence of the
Hamiltonian in a much quicker though less pedestrian way than Feynman's. Note, that these hidden
assumptions are not necessary for classical systems, since one can then argue with the local 
expression for the Poisson bracket. Therefore Feynman's proof of the existence of a Hamiltonian is
a much less trivial statement in the classical case.

\noindent
In this paper we generalize Feynman's argument to arbitrary configuration spaces $\CQ$.
For this purpose, we replace the algebra generated by the $x_i$ by the algebra
$C^\infty_0(\CQ)$ of smooth functions which vanish at infinity. Our axioms for
nonrelativistic quantum mechanics could then be completely formulated in the language of
(commutative) operator algebras. However, this might seem to hide its essential physical
features. We therefore adopt a more geometric -- sometimes  sloppy -- language here.
The precise operator-algebraic formulation will be given in the sequel \cite{us3} to the present article. 

\noindent Our formalism is completely background free, in the sense that it does
not use a metric on the configuration space, but rather it reconstructs the metric from the 
given time evolution of the quantum mechanical system.

\noindent
As a side result we obtain a formulation  of quantum mechanics, that can be generalized
to many (though not all) noncommutative configuration spaces. This will be explained in \cite{us3}.

\noindent
We hope, that this work clarifies the appearance of geometrical entities describing all the
consistent interactions -- we have tried to find geodesics not only in spacetime , but also
in our proofs.

\section{Axioms for nonrelativistic quantum mechanics}
Let $\CQ$ be any smooth, orientable manifold. Following Feynman's original idea, we shall try to
set up quantum mechanics for a scalar particle over the configuration space $\CQ$ starting from as
few empirical facts as possible. In fact, we shall need less assumptions than Feynman. \\
In general, there will be no {\em globally} well-defined coordinates on $\CQ$ which could be
represented as operators on an appropriate Hilbert space.
We shall therefore work with the algebra $\CA= C^\infty_0(\CQ)$ of smooth, real-valued
functions over $\CQ$ which vanish at infinity. All the physical observables will then be
constructed from $\CA$, or, more precisely, from its representations on the physical Hilbert space
$\CH$ :\\
It is well-known (see[] ), that only the Hilbert spaces  $\CH=L^2(\CQ,E)$, i.e.
the spaces of square integrable sections of (flat\footnote{for electrically neutral particles})
complex line bundles 
\[  E\stackrel{\pi}{\longmapsto} \CQ  ,        \]
can serve as candidates for quantum mechanical models. From the algebraic point of view,
using the algebra $\CA$, the representations ( by pointwise multiplication) as {\em
selfadjoint, bounded operators} on these spaces could (essentially) be characterized  by two
conditions:
\begin{itemize}
\item They are norm-preserving and {\em faithful}. Intuitively, this requires that the particle can (in
      principle)
      be localized (in the sense of generalized eigenstates of $\CA$) in {\em every} point of
      $\CQ$. 
\item  For scalar particles there are no
       observables which commute with all the position operators. Thus every operator that commutes with
       all the (real, smooth) functions must either be a (complex) function on $\CQ$ or a complex multiple
       of the identity (which will not be in $\CA$ if $\CQ$ is noncompact, since it does not vanish at
       infinity). Denoting the complexification of $\CA$ by $\CA_\C$ and the commutant by
       $\CA'$,  we shall therefore require:
       \[  \overline{( \CA_\C + \C\ei ) } = \CA'  .\]   
\end{itemize}
We still need time:  We shall work in Heisenberg picture. Hence it is assumed that the operators representing
$a\in \CA$ depend smoothly (in the strong sense) on the parameter $t\in \R$. One may view this as a family
of representations of $\CA$, parametrized by the time $t$. To keep the notation short, we shall denote the
representation at time $t$ by $\CA_t$, with elements $a_t,b_t,\ldots \in \CA_t$. 
Thus $a_t\in \CA_t$ is the representation of a function $a$ on the
Hilbert space of  square integrable solutions to the Schroedinger equation with values in the bundle $E$.\\
(Using local coordinates $q_i$ and their time evolution $q_i(t)$ in the Heisenberg picture
one may think of as $a_t = a(q_i(t))$, which might be a more familiar notation.) We shall also assume that
the time evolution is norm preserving:
\[ \|a_t\| = \|a \| \quad\qquad\qquad \forall a_t\in \CA_t \qquad\forall t\in\R .   \]
Note that, due to the
smearing of wave functions, representations $\CA_{t_1},\CA_{t_2}$ at  different times
$t_1 < t_2$ do not commute: \\ They will have different sets of (generalized) eigenvectors since
a wavefunction localized in point $q\in \CQ$ at a time $t_1$ will no longer be localized in some point
of $\CQ$ at the later time $t_2$.
\\

The requirement of smoothness then (essentially) demands the existence of
\[  \dot{a}_t \,\,\stackrel{def}{=}\,\, \left. \frac{d a_\tau}{d\tau}  \right|_{\tau=t}    
\]
as {\em selfadjoint, bounded} operators on $\CH$ for all $t\in\R$ and all $a_t\in \CA_t$. The
subset of the bounded operators ${\cal B}(\CH)$ of all $\dot{a}_t $ is denoted by
$\dot{\CA}_t$. \\ The only further
empirical input we shall require is the analogue of the canonical uncertainty relation over $\CQ$:
The uncertainty in a measurement of velocities and positions  at equal times shall be nontrivial
and independendent of the velocities, i.e. it is supposed to be a function of the positions
alone.\\ 
We should stress that, unlike Feynman, we do not require  a particular form of the commutator
of functions and velocities. We only demand it to be nontrivial and a function of the positions
only !

Let us (sloppily) summarize:\\

\begin{defn}
Let $\CA=C^\infty_0(\CQ)$. We call a family $\left( \CA_t, \,\, t\in \R \right)$ of 
unitary representations of $\CA$  {\bf scalar quantum mechanics over $\CQ$}, if the
following conditions are satisfied:
\begin{enumerate}
\item {\bf Localizability:} The representations $\CA_t$ are isomorphic to the representations
       of $C^\infty_0(\CQ) $ on the space $\CH= L^2(\CQ, E)$, where $E\stackrel{\pi}{\mapsto}\CQ$
        is some complex line bundle over $\CQ$. 
\item {\bf Smoothness:} The time evolution, i.e. the
      dependence of the representations
       on the parameter $t$, is smooth (with respect to the strong topology) and:
        \[ i[ \CA_t, \dot{\CA}_t ] \subset \CA_t \qquad \qquad \forall t\in \R     \]     
\item {\bf Nontriviality and positivity:}
\[ -i[a_t , \dot{a}_t] \geq 0 \qquad\qquad\qquad  \forall a_t\in \CA_t . \]
 If  \[ [a_t,\dot{a}_t] = 0 \qquad\qquad\qquad \]for some $a_t\in\CA_t$,
then
      $\dot{a}_t= 0$. 
\end{enumerate}
\end{defn} 

\noindent 
We still need to comment on the "nontriviality"-assumption: \\
For $\CQ = \R^n$, when there do exist global coordinates (which are represented on $\CH$), then the
existence of a nontrivial function with $[a_t,\dot{a}_t]=0$ would be
equivalent to the statement, that there exists a coordinate system $q_i$ such that
\[ [q_k, \dot{q}_k ]  = 0  \] for some $k$. Thus the $q_k$-direction is "unquantized".   
In the general case, one might similarly conclude the existence of local unquantized coordinates.\\
Due to this nontriviality axiom it is clear, that $ -i[a_t , \dot{a}_t]$ is either
positive or negative for all $a_t$. The sign can be changed by a time reversal $t\to
-t$ , so that one can always achieve positivity. As we shall see this convention corresponds to a 
Hamiltonian that is bounded from below. \\

\noindent
It should be emphasized that these axioms do not use any metrical structure on $\CQ$. 
As we shall see, one can reconstruct from any scalar quantum mechanics the metric on $\CQ$ to which it 
corresponds.
 
\section{The most general Hamiltonian....}
Let us now assume that we are given a family of representations $\CA_t$ obyeing the above conditions.\\
First of all, it is implied therein that all the representations $\CA_t$ are unitarily
equivalent. Hence, there exists a strongly continuos semi-group of unitaries $U(t_1, t_2)$
which connects these representations,
\[   \CA_{t_2} = U(t_2,t_1) \,\CA_{t_1} \,U(t_1,t_2) .    \]
A well-known theorem of Stone then ensures the existence of generators $H(t)$ for this
semigroup.\\ All that would have also followed from slightly weaker assumptions, like the algebraic version
of the Localizability axiom, which we suggested above.   The precise -- but lenghty -- formulation of the
smoothness condition will render these statements completely rigorous. We shall then also give the easy proof
of the following\\
\begin{corollary}
Let $(\CH , \CA_t)$ be a scalar quantum mechanics over $\CQ$. Then: 
\begin{enumerate}
\item \[ \overline{\CA_t} = \left(\CA_t \right)_+ ' \qquad 
         \qquad\qquad \qquad \qquad\forall t\in\R      . \]
\item There exists a function  $H(t)$ from $\R$ to the set of  formally selfadjoint
(generally unbounded) operators with a dense domain $\CH_\infty \subset \CH$, such that 
     \[  \dot{a}_t = i \,[a_t, H(t) ] \qquad \qquad \forall t\in \R, \qquad \forall a_t\in \CA \]
\end{enumerate}
\end{corollary}\noindent
\\
That is more or less put in via the axioms. However, we have not yet  used "the cure": \\
We denote the space of real-valued one-forms over $\CQ$ by $\Omega^1\CA$. A {\bf metric} on $\CQ$
is then a symmetric,  bilinear (over $\CA$) map 
\[ g: \quad  \Omega^1\CA \otimes_{\CA} \Omega^1\CA \,\, \longrightarrow \CA .\]
Bilinearity over $\CA$ means that one has $ g(a_1\, {\rm d} b_1\, ,\, a_2\, {\rm d}b_2) = a_1a_2\, 
g({\rm d} b_1\, , \, {\rm d} b_2)$. \\
It is nondegenerate, if from
$ g(\omega \,,\, \alpha) = 0$ for some $\alpha$ and all $\omega$, it follows that  
$\alpha = 0$.
A (positive definite)  {\bf Riemannian metric} on $\CQ$ is a nondegenerate metric such that 
$g(\omega\, ,\, \omega)\geq 0$ for all $\omega\in \Omega^1\CA$. There do always exist Riemannian
metrics on $\CQ$.\\
\begin{lemma}
There exists for every $t\in \R$ a  Riemannian metric $g_t$ on $\CQ$, such that:
\[  [a_t , \dot{b}_t] = i\, g_t({\rm d}a_t \,,\, {\rm d}b_t) \qquad\qquad \qquad \forall a_t\,
,\, b_t\in \CA_t . \]
$g_t$ depends smoothly on $t$. 
\end{lemma}
\proof \\
We only sketch the proof.
To simplify the notation, we shall suppress the subscript $t$ for a while. Define then
\[  g_t(c_1{\rm d}a \,,\, c_2{\rm d}b) \stackrel{def}{=} -i\, c_1 c_2 \,[a , \dot{b}] 
   \qquad \qquad  a ,b, c_1,c_2\in \CA_t .\]
Note that $g_t$ is well defined, i.e. compatible with the {\em Leibniz rule} for the differential
d: \begin{itemize}
\item The Leibniz rule in $a$ follows from  
\[ [a_1 a_2 \, , \, \dot{b} ] = a_1[a_2 \, , \, \dot{b} ] + a_2[a_1 \, , \, \dot{b} ]   .\]
\item The Leibniz rule in $b$ follows from \[\dot{(b_1b_2)} =  \dot{b}_1 b_2 + b_1\dot{b}_2 .\]
\end{itemize}
(Here we have used that $[a \, , \, \dot{b} ] \in \CA\subset \CA'$.)\\
Thus, we have constructed a bilinear map $g_t \, :\, \Omega^1\CA \otimes_{\CA} \Omega^1\CA \,\,
\longrightarrow \CA $ .\\
It is symmetric, since
\[    [a\,,\,b] = 0 \quad \Rightarrow \quad [\dot{a}\, ,\,b ] + [a \,,\, \dot{b}] = 0 .\]
Due to the third axiom of nontriviality and positivity it is also positive
 definite and nondegenerate.
That should be obvious, as well as the smoothness in $t$.
\hfill\endproof \\
\begin{remark}
Note, that for simplicity we have chosen to set $m=1$, thereby fixing a length scale.
A generic $m$ would, of course, appear in the commutator -- replacing $g_t$ by $\frac{1}{m}g_t$.
\end{remark}

\noindent
A smooth {\bf vector field} over $\CQ $ ia a linear map $X\, :\, \CA\to\CA$ obeying the Leibniz
rule. Obviously, the map $a_t\mapsto -i[a_t,\dot{b_t}]$ defines such a vector field for every
$b_t$. This is the canonical vector field $X_b$ on the Riemannian manifold $(\CQ, g_t)$ associated
to $b_t$ It is defined by $X_{b_t}(a_t) = g_t({\rm d} b_t \, ,\, {\rm d} a_t)$.\\ 
If the bundle $E$ is trivial, i.e. if $\CH$ is the space of square integrable complex functions,
then every smooth  vector field $X$ can be  represented on (a dense subspace $\CH_\infty$ of) $\CH$
in the obvious way if one thinks of it as a differential operator of first order. By the
Leibniz rule, one then has:
\[  [iX, a_t] \psi =  iX(a_t)\psi  \qquad\qquad \forall a_t\in \CA_t\qquad 
    \forall \psi\in \CH_\infty. \]
However, this representation is not unique, and, more importantly, it is not well-defined
if the bundle $E$ is nontrivial.\\
In the latter case one still can define a map $\nabla$ which assigns to every smooth vector field
$X$ an operator $\nabla_X\, :\, \CH_\infty\to\CH_\infty$, called the {\bf covariant derivative
along $X$}, such that 
\bea
\nabla_X^* & = & \nabla_X  \nonumber \\
\nabla_X \left(a_t \psi\right) & = & iX(a_t)\psi + a_t \nabla_X\psi \label{leir} \\
\nabla_{a_t X_1 + b_t X_2} & = & a_t\nabla_{X_1}+ b_t \nabla_{X_2} \label{linr}.
\eea
Covariant derivatives do exist on every complex bundle. They are, however, not uniquely defined
by the above conditions: Let $\nabla^{(1)}, \nabla^{(2)}$ be two such covariant derivatives.
It then immediately follows from (\ref{leir}) that 
\[    \alpha(X)  \stackrel{def}{=} \nabla^{(1)}_X -\nabla^{(2)}_X  \]
commutes with all functions $a_t$ and hence $\alpha(X)\in \CA$ for all smooth vector fields $X$.
Due to (\ref{linr}) the map
$\alpha$ is linear, i.e. there exists a one-form $A\in \Omega^1\CA$ such that:
\[  \alpha(X) =  \langle A,X \rangle.\]
Thus the covariant derivatives on the line bundle are in one-to-one correspondence with real one
forms $A$. Fixing $\nabla$, and denoting the covariant derivative along
the canonical vector fields $X_b$ by $\nabla_b$, one then easily computes (suppressing $t$):
\[ [a,\nabla_b] = -iX_b(a)  = [a, \dot{b}]  \]
Hence $\nabla_b$ and $\dot{b}$ can at most differ by a function $\beta(b)$. From the
definitions of $X_b$ and $\nabla$ also follows the Leibniz rule
\[ \nabla_{(bc)}= b\nabla_c + c\nabla_b \qquad\qquad \qquad \forall b,c \in CA  \]
which holds for the time derivative as well, and hence $\beta$ is a vector field. Using the metric
$g_t$ we therefore obtain another real one-form $B$, defined as $\beta(b_t) = g_t(B,{\rm d}b_t)$.
$B$ can then be absorbed by a redefinition of the connection $A$:\\
Indeed, one checks that for any two connections differing by a one-form $A$, one
has: 
\[  \nabla^{(1)}_b -\nabla^{(2)}_b  = g(A,{\rm d} b)   .\] 
So finally we arrive at \\
\begin{lemma}
For every $t$ there exists a unique covariant derivative $\nabla (A_t,g_t)$ on $\CH$ such that
\[  \dot{b_t} = \nabla_{b_t}(A_t,g_t) \qquad\qquad\qquad  \forall b_t\in \CA_t . \]
\end{lemma} 

\begin{remark}
In the language of noncommutative geometry the space of sections of a vector bundle
is characterized as a projective module over $\CA$. Thus, there exists an integer $n$
and a selfadjoint $n\times n$-matrix $p=p^2$, i.e. a projector,  with entries from $\CA$
such that 
\[ \CH_\infty = p\CA^n  .\]
Since one can represent all vector fields $X$ on $\CA^n$,  one then
immediately infers the existence of a covariant derivative $\nabla_X$ on $\CH_\infty$:
\[  \nabla_X = p \left( X\otimes \ei_n \right) p = pX + \underbrace{p [X\otimes\ei ,  p]}_{=:
p\langle A,X\rangle} .\]  If $p$ is nontrivial, then the connection one-form $A$ is obviously
needed  to make $\nabla_X$ well defined. \\
Recall also that covariant derivatives are in one-to-one correspondence with
connections (therefore also denoted $\nabla $) on $\CH_\infty$, i.e. linear maps from
$\CH_\infty$ to $\CH_\infty \otimes\Omega^1\CA$ obyeing the Leibniz rule in the sense of
(\ref{leir}). Explicitly the isomorphism is given as:
\[ \langle \nabla \psi , X \rangle = \nabla_X \psi  \qquad\qquad \forall \psi\in
\CH_\infty .\]
\end{remark}

\noindent
With this result at hand, let us now return to the Hamiltonian:\\
The last ingredient from differential geometry we shall need is the covariant
Laplacian $\Delta (A_t, g_t)$, defined via the covariant derivatives $\nabla (A_t,g_t)$.
Using {\em local} coordiantes $q^j$ and the local basis $\frac{\partial}{\partial q^j}$ of the
tangent space over $\CQ$, it is given as
\[   \Delta (A_t,g_t) = \oh\sum_{ij} g_t^{ij} \left(
\frac{-i\partial}{\partial q^i} - (A_t)_i \right) 
\left(\frac{-i\partial}{\partial q^j} -(A_t)_j\right) .\]
It is, of course, a {\em globally} well-defined operator.
Note, that $\nabla_\frac{\partial}{\partial q^j}(A_t,g_t) = \left(\frac{-i\partial}{\partial q^j}
-(A_t)_j\right)$. Thus the covariant Laplacian is a well-defined, self-adjoint operator
on $\CH_\infty$. Using the above local form, one then  verifies
\[ \dot{b}_t =  i[\Delta(A_t,g_t),b_t] = i[H,b] . \]
Thus, the Hamiltonian $H$ of the system can differ from $\Delta(A_t,g_t)$ only by a function, and we obtain
our main result:\\
\begin{thm}
There exists (for all $t$) a unique Riemannian metric $g_t$ on $\CQ$, a unique covariant
derivative $\nabla(A_t,g_t)$ on the complex line bundle $E$ over $\CQ$ and a 
(differentiable) function $\varphi_t$ such that
\[  H(t) = \Delta(A_t,g_t) + \varphi_t.  \] 
\end{thm}

\begin{remark}
In contrast to Feynman's argument,  we did not assume that the equations of motion are of second
order, i.e. Newtons second law, but rather derived it. In fact, the "cure" 
\[ [a_t,[b_t,H(t)]] \in \CA_t \qquad\qquad\qquad \forall a_t,b_t \in \CA_t \]
implies that the Hamiltonian is a differential operator of second order. \\
Note that $H$ is bounded from below (if $\varphi$ is)  due to the positivity of the metric.
\end{remark}
\section{...and spacetime}
It is quite instructive to study the physical interpretation of our result.
In order to proceed step by step, let's first assume that the bundle $E$ is trivial.
In that case, it is possible, that the connection one-form $A$ vanishes.
The classical analogue of the above Hamiltonian $H$ then reads in local Darboux-coordinates
$(\vec{q},\vec{p})$ on $T^*\CQ$:
\bea H(t) = \oh \sum_{ij} g_t^{ij}(\vec{q}) \, p_ip_j . \label{geo} \eea
If the metric $g_t = g$ does not depend explicitly on $t$, then the Hamiltonian
equations of motion for this system describe the geodesics on $(\CQ,g)$:
\[ \ddot{q}^k +\Gamma_{ij}^k \dot{q}^i \dot{q}^j = 0  \]
where $\Gamma^k_{ij} = \oh \sum_l\, g^{kl} (\partial_ig_{lj}+\partial_j g_{li}-\partial_lg_{ij})$
are the usual Christoffel symbols describing a torsion-free Levi-Civita connection.\\

\noindent
For a generic time-dependent metric $g_t$ the situation is, however, more complicated: 
\\ Let's introduce
the extended configuration space $\CM = \CQ \times \R$, where $\R$ refers to the time $t$.
The metric $g_t$ is then lifted to the Lorentzian metric
\bea {\bf g} = \left(  \begin{array}{cc}  -1 & 0 \\0 & g_t \end{array}\right)\eea
on $\CM$, so that we describe a {\bf globally hyperbolic spacetime $(\CM,{\bf g})$}
in special coordinates, where the time is chosen orthonormal to the spacelike hypersurfaces
$\CQ_t \sim\CQ$.

\begin{remark}
One can then view the representations $\CA_t$ of our quantum mechanical models as
being representations of the functions $C^\infty_0(\CQ_t)$ on the time slices $\CQ_t$.
The smoothness of the time evolution ensures that the time slices are glued together
smoothly, thus describing the full spacetime $\CM$. This is the approach to an algebraic
description of globally hyperbolic space times developed in \cite{us2}. 
\end{remark}

\noindent
The
additional fields $A$ and $\varphi$ will then, of course, describe the electromagnetic potentials on $\CM$.
If one introduces the connection one-form
\[ {\bf A} = A + \varphi {\rm d}t  \]
on $\CM$, then obviously the equations of motion only involve its curvature (field strength)
${\bf F} = {\rm d}{\bf A}$. From the existence of the potential ${\bf A}$ then immediately follow
the homogeneous Maxwell equations $${\rm d}{\bf F} = 0.$$
The inhomogeneous equations $*{\rm d}*{\bf F} = {\bf j}$ will then serve as the definition of the
{\em external} sources ${\bf j}$. (Similarly we could use Einstein's equations to define
the external sources ${\bf T}^{\mu\nu}$ of the gravitational field.)\\ 
However, let's neglect  ${\bf A}$ again, because we are only interested in  the effect of the metric
$\tilde{g}$ in this paragraph.

\noindent
Since we are now working with a relativistic system on $\CM$, we should switch to the Lagrangean
formalism. The Lagrangean $$L= {\bf g}_{\mu\nu} \, \dot{q}^\mu\dot{q}^\nu  $$
then obviously leads to the geodesic equations on $(\CM,{\bf g})$ for the spatial components $k$.\\
But in the special coordinates, where $q^0=t$ and ${\bf g}_{00}=-1$, ${\bf g}_{0i}=0$,
it reduces to $$L= \sum\limits_{ij}  g_t^{ij}(\vec{q})  \,\dot{q}^i\dot{q}^j -1 ,$$ thus being
equivalent to (\ref{geo}). \\
However, the zeroth component of the equations of motion for this Lagrangean
\[ \underbrace{\ddot{q}^0}_{=0} +\oh\frac{\partial g_{ij}}{\partial t} \dot{q}^i \dot{q}^j = 0  \]
are only consistent if the metric $g$ does not depend on time.

\noindent
In conclusion, the Hamiltonian (\ref{geo})  does only describe the geodesic motion
on the spacetime $(\CM,{\bf g})$ if the metric ${\bf g}$ is static. In order to describe
a generic (time dependent) gravitational field one would therefore have to consider relativistic
quantum mechanics. Nevertheless the nonrelativistic case is sufficient to describe all Lorentzian metrics
${\bf g}$ on an arbitrary globally hyperbolic spacetime $\CM$ .\\

\begin{remark}
Note that the sign of ${\bf g}^{00}$ does not play a role for the
homogeneous Maxwell equations. (It only enters the inhomogeneous equations via the
Hodge-$*$-operation.)
Therefore, we might have also used an euclidean metric on $\CM$. Our choice has been motivated by
physics only.\\
This then answers Dysons question, why only relativistic covariant interactions are compatible
with the nonrelativistic uncertainty relation, in a somewhat unsatisfactory way: 
The only thing we can safely conclude from our analysis is that the relativistically
covariant electromagnetic interacions are consistent with the uncertainty relation -- in a special
frame, of course.
\end{remark}

\noindent
In the local Darboux coordinates the construction of the most general Hamiltonian
is actually quite easy and can be performed also in the classical context.
In that case, one obtains partial differential equations, which are easily integrated
locally. One then also checks that the local expressions can be glued together
appropriately to give a global solution.\\
In the quantum context, the global aspects of the geometry of $\CQ$ -- and the differential forms
entering the Hamiltonian -- play a much more important role, however. On the one hand, the Hilbert
space
$\CH$ might be the space of square integrable sections of a nontrivial bundle. That is
taken into account in our framework in a very explicit way. \\
If one wants to describe electrically neutral particles, one will choose the connection one-form
${\bf A}$ flat, i.e. such that its curvature ${\bf F}$ vanishes. That is, of course, only possible on
flat bundles. \\
If $\CQ$ is simply connected, then this flatness requirement defines a unique (and thus
canonical) connection on $\CM$. However, if the fundamental group $\pi_1(\CQ)$ is nontrivial,
then there might exist a flat connection associated with every unitary one-dimensional representation of
$\pi_1(\CQ)$ even for the trivial line bundle. These representations will be realized as holonomies
of ${\bf A}$ around the elements of $\pi_1(\CQ)$.\\
To give an example, we consider the space $\CQ= S^1\times \R$, i.e. a cylinder.
Note that all complex line bundles on the cylinder are trivial. Nevertheless there do
exist infinitely many nonequivalent flat connections. Intuitively this might be seen as
follows:  Imagine the cylinder being imbedded in $\R^3$, with a solenoid in its interior.
Then the magnetic field strength on the cylinder will vanish. However, a quantum mechanical
(charged) particle on the cylinder will nevertheless feel the (quantized) magnetic flux through 
the interior
via the Aharonov-Bohm effect. \\
The incorporation of such topological effects has been one of the motivations for the present 
article.
There are many physical applications of such phenomena. For instance, the appearance of anyonic
statistics in $2+1$ dimensions is directly related to the above example of the cylinder
(which is the configuration space for the relative motion of two particles in two dimensions).
For two particles in $3+1$ dimensions, when the fundamental group is $\Z_2$, the 
Fermi-Bose-alternative can be recovered by a similar argument. However, these examples show
that we still need further axioms in order to describe {\em only} realistic quantum systems: 
The present set of axioms does not exclude the wrong connection of spin and statistics.
A systematic investigation of the Spin-Statistics-Theorem in nonrelativistic
quantum mechanics is currently in progress \cite{SpinStat}.

\section{Outlook}
This article resulted from an attempt to understand the geometry behind "Feynman's proof
of the Maxwell equations" and the necessary assumptions it needs. That then lead us to a set of
axioms for nonrelativistic (scalar) quantum mechanics, which are completely formulated in the
language of algebras of observables. If one replaces the concrete algebra $C^\infty_0(\CQ)$
by an arbitrary commutative pre-$C^*$-algebra (fullfilling certain ``smoothness'' requirements),
then these axioms would, in fact, be formulated without any reference to a fixed geometric 
background. Rather, our result would provide a way to reconstruct the geometry of spacetime
from the basic commutation relations of the theory. Of course, for the nonrelativistic theory we 
have been working with, that is not such a spectacular new result. The real challenge of 
finding a background free formulation of quantum theories is the requirement of
{\em locality} in relativistic theories, i.e. the requirement that local observables at 
{\em spacelike} separated points commute. It is therefore tempting to generalize our
approach to relativistic particles.\\
We tried
to find an argument as simple as possible, hoping that this facilitates a generalization to 
other geometries, such as Lorentzian manifolds $\CQ$.
In the sequel to this paper we shall, however, first examine examples for noncommutative geometries.
Besides the fashionable Moyal-deformed $\R^n$,  we also consider the noncommutative algebra
$C^\infty_0(\CQ)\otimes M_k(\C)$ of matrix-valued functions on $\CQ$. Thereby we 
reproduce non-abelian Yang-Mills theories, but also the Pauli equation, i.e. the coupling of
the spin to the motion and external magnetic fields. \\
As expected, this latter result requires some additional input that appears rather unnatural.
The correct treatment of the spin degrees of freedom should come out naturally only in a
relativistic theory. In fact, our main motivation for this project has been the generalization of
Feynman's argument to relativistic particles, which is still an open problem. 
We shall present a physically motivated solution in []. Surprisingly, the axioms
for "relativistic quantum mechanics for spin-$\oh$" that we found are very similar to Alain Connes' axioms
for noncommutative spin manifolds \cite{Preis}. \\

\vspace*{2cm}
\noindent
\begin{center}
{\bf \large Acknowledgements}
\end{center}

\noindent
I would like to thank T.Kopf, A.Sitarz and A.Sergyeyev for valuable discussions
on the subject. It is a pleasure to thank the A.von-Humboldt-foundation for its support.

\end{document}